# CROWD-SHIPPING SERVICES FOR LAST MILE DELIVERY: ANALYSIS FROM SURVEY DATA IN TWO COUNTRIES

**Tho V. Le**
Ph.D. Student
Lyles School of Civil Engineering
Purdue University
550 Stadium Mall Drive, West Lafayette, IN 47907, US
Tel: +1 765 586 2836; Email: le39@purdue.edu

**Satish V. Ukkusuri, Corresponding Author**
Professor
Lyles School of Civil Engineering
Purdue University
550 Stadium Mall Drive, West Lafayette, IN 47907, US
Tel: +1 765 494-2296; Fax: +1 765 494 0395; Email: sukkusur@purdue.edu

Word count:  5,052 words text + 7 tables/figures x 250 words = 6,802 words

Submission Date August 1st, 2017

## ABSTRACT

The e-commerce boom has led to overwhelming demand for personalized delivery services. Accordingly, various start-ups and tech companies provide crowd-shipping services that aim to be more efficient and effective than traditional logistics options. These services are fueled by technological innovation, improved internet infrastructure, and increased smartphone use. However, the field of on-demand delivery faces several challenges, including specified pickup and delivery times and locations. Therefore, market demand and prospective crowd-shipper supply must be well understood to ensure industry success. This research analyzes current and future shipping behaviors, as well as potential employees' willingness to work (WTW) as crowd-shippers. Revealed and stated preference survey questionnaires were designed. The surveys were implemented in Vietnam and the US. This descriptive study makes use of the survey data sets to understand the behavior of requesters and potential crowd-shippers in the logistics market and assumes that crowd-sourced delivery is available. The results show requesters' various behaviors and expectations as well as prospective crowd-shippers' WTW in the two countries. The results can be used to recruit potential crowd-shippers and create business strategies that match requesters' and potential crowd-shippers' expectations.

## INTRODUCTION

The *sharing economy* is a popular term, which has been widely used in recent years. "It includes the shared creation, production, distribution, trade and consumption of goods and services by different people and organizations" (*1*). In transportation, there are two main forms of sharing, i.e., of passengers (real-time ridesharing and taxi share) and of freight (app-based services like crowd-shipping).

*E-commerce* has sharply increased all over the world, and it has already changed the shopping habits of people and created new challenges for logistics providers (*2*). The top 10 national economies have seen double digit growth figures in the e-commerce market (*3*). In addition, Hayashi et al. (2014) pointed out that not only the price of merchandise but also delivery fees are important to consumers (*4*). According to a survey presented in *E-commerce*, delivery, price, convenience, and speed of delivery are important features for repeated online purchases (*5*). Accordingly, delivery services significantly influence the success of online shopping (*4*).

Numerous crowd-shipping firms have been established worldwide. Crowd-shipping, or crowdsourced delivery, is "an app-based platform that connects the individual wanting to ship a packet with an individual willing to carry the shipment in the first or last mile logistics of urban areas. A key distinction of the courier as discussed in this research is that this is *not necessarily an additional trip but a trip that leverages the typical travel patterns of the courier*. The selected courier may be the closest to the delivery route, offer the cheapest delivery fee, or have the best reputation in the system" (*6*). Crowd-shipping companies provide services for international, national, regional, and urban deliveries. In fact, the delivery and shipping companies founded between January 2015 and September 2016 have received the highest investment from entrepreneurs (*7*).

Clearly, knowledge of the demand and supply sides benefits crowdsourced delivery companies. Demand side details address the needs and expectations of customers (requesters). In addition, understanding the incentives of people to participate in a crowd-shipping system is worthwhile for companies to establish their workforce. Moreover, expertise in the logistics market is critical for building sustainable business models. Pricing policies and incentives for drivers play a crucial role in these business models, in addition to an attractive and user-friendly app. Therefore, the goals of this research are to provide insights into the behavior of requesters and potential crowd-shippers in a logistics market with a given crowd-shipping availability. This research is needed to understand the impacts of this service not only for crowd-shipping companies but also for city agencies.

This paper is organized as follows. Section 1 introduces the research questions of motivations for this paper. In section 2, we identify the gaps related to the research on crowd-shipping. Section 3 discusses our questionnaire design, which includes the revealed preference (RP) and stated preference (SP) questions. The survey implementation is also included in section 3. Section 4 illustrates the descriptive analysis and insights of our survey data from Vietnam and the US. Further suggestions are offered in the discussions and recommendations section, which is followed by the conclusion.

## IDENTIFYING THE GAPS

Various delivery services specialize in grocery, food delivery (e.g. Instacart, Postmates, and UberEats), book delivery (e.g. Piggybaggy), or all non-hazardous items (e.g. Doordash, OrderUp, and Roadie). These novelty services are expected to transform the way people ship and offer various advantages to consumers, retailers, and society at large (*6, 8-10*).

Given that crowd-shipping is an emerging market, there are few research studies on crowd-shipping to date. A couple of publications examine the supply side of crowd-shipping services. Miller et al. (2017) developed models to better understand travelers WTW as crowd-shippers. Interestingly, only 43 of 143 respondents (about 30%) were willing to work as crowd-shippers. The findings reveal that travelers have WTW values much higher than the conventional willingness to pay (WTP). Service attributes, socio-demographic characteristics, and attitudinal variables remarkably influenced respondents' WTW (*11*). Furthermore, Le and Ukkusuri (2018) developed discrete-continuous models to understand various insights related to WTW and travel time tolerance. Potential crowd-shippers' payment expectations were found to be reasonable and concurrent with the value of time in literature. Socio-demographic characteristics significantly affected respondents' WTW (*12*). Paloheimo et al. (2016) studied a pilot crowdsourced delivery service for a library in Finland and identified various motivations for participation, including "try out something new," "make life easier for me," "support public services," and "support the environment" (*13*). Other papers presented brief results related to crowd-shipping preferences (e.g. 32 of 50 respondents (64%) are willing to work as crowd-shippers) (*14*), or factors affecting friends crowdsourcing delivery for friends through social networks (e.g. ages, incomes, genders, and extra time demands) (*15*).

Several researchers have studied the demand side of crowd-shipping services as well. Punel and Stathopoulos (2017) modeled the openness of consumers to crowd-shipping services. Travelers' behaviors and preferences for crowdsourced deliveries were significant based on the travel distance as well as the following aspects: speed, real-time tracking (local delivery), service options, and driver experience (medium and long distance) (*16*). Interestingly, the percentage of respondents who had already used crowd-shipping services was 7% (*16*) and 12% (as quoted in (*17*)). Briffaz and Darvey (2016) reported that 74% of respondents (37 of 50) were willing to use crowd-shipping services (*14*). Research by Acquity Group LLC found that 75% of respondents were willing to use delivery services from third parties (as mentioned in (*17*)).

These preliminary studies have several limitations, and more studies are necessary to further analyze the crowd-shipping field. Accordingly, the objectives of this paper are as follows:

- Factors related to *crowd-shippers*' behavior:
  - Question group (QG) 1: Who is willing to work as a crowd-shipper? Are there any specific socio-demographic characteristics associated with those people? What factors drive them? Do they have any preference for shipment types or requesters?
  - QG2: How much do crowd-shippers expect to be paid (ETP)? What is the sensitivity of crowd-shipper incentives and WTW?
  - QG3: What is the maximum distance or travel time a crowd-shipper would accept to divert their route to pick up and deliver packages? What factors influence these decisions?
- Factors related to *requesters*' behavior:
  - QG4: What factors influence requesters' selection of couriers for different type of products?
  - QG5: How much are senders willing to pay for last-mile delivery? What is the nature of the relationships between socio-demographic characteristics, products, and WTP?

In addition, we are interested in understanding the underlying behavior rules for the demand and supply generations of different cultural contexts. Surveys were conducted in the US (America) and Vietnam (Asia). The next section presents our questionnaire design, which addresses the aforementioned objectives.

## QUESTIONNAIRE DESIGN

Crowd-shipping is a modern concept for many people (43% have never heard about it (*16*)) so the questionnaire should be easy to understand and well-designed to capture the necessary information. This survey aims to investigate the behavior of stakeholders in the logistics market given the availability of crowd-shipping services. The questionnaire consists of three main parts. Surveys were designed in Qualtrics, an online survey platform, and surveys links were distributed. The following sections discuss the RP sections of parts IA and IIA, and the SP sections of parts IB and IIB. The survey implementation methods are featured in the last section.

### Revealed preference sections

In part IA, featuring the sender's behavior, questions were designed to ask respondents either 1) their most recent shipping activity experience with carriers, or 2) their satisfaction with the delivery service from their most recent e-commerce purchase. Questions related to the commodity value, delivery carrier, shipping cost, delivery time, and delivery time satisfaction were included. In addition, questions related to the satisfaction with carriers' other services—tracking and tracing items online, electronic delivery notifications, and pickup/drop-off time windows and locations—were featured as well. Tipping behavior was the focus of a question in the US survey.

    In the part IIA, regarding the courier's behavior, questions were designed to obtain the couriers' history and their perceptions. Respondents were asked: 1) whether they have transported freight for someone else, and 2) in the future, if they had a chance to transport freight for somebody else on their route or close to their route of travel, are they willing to work for some incentives? If so, they were asked in which situations they were willing to do the work. The maximum diversions (both for time and distance) the respondent would accept for picking up and dropping off a package included. Respondents were also asked their ETP once they began work as a crowd-shipper.

### Stated preference sections: Attributes and levels of service

Regarding the sender's behavior, part IB is designed to understand the behavior of selecting couriers for each product shipping category (e.g. shipping beverages/dried foods, apparel, and personal health/medicine). These SP questions were created with exactly the same attributes as the RP portion with the intent to use a combined model (RP+SP) for later data analysis. The choice set includes four alternatives with the attributes and levels of services presented in Table 1(a). In addition to the two traditional attributes (i.e., shipping cost and delivery time) indicated as the main factors of delivery services (*18-20*), we introduce attributes based on the major differences of crowd-shipping services and traditional logistics services (Table 1 (a)).

    In addition to the SP questions, other questions related to the preference of delivery time windows, concerns about delivery by a crowd-shipper, and preferences on the delivery mode were also asked.

### Table 1 (a). Attributes to understand the requesters' behavior for selecting couriers

| q | Attribute | $l_q$ | Level |
|---|-----------|-------|-------|
| 1. | Shipping cost (values of shipping cost in the US are outside the parenthesis, in Vietnam are inside the parenthesis) | 1 | $14 (20,000 VND) |
| | | 2 | $18 (30,000 VND) |
| | | 3 | $22 (35,000 VND) |
| | | 4 | $26 (40,000 VND) |
| 2. | Delivery time | 1 | 1.5h |
| | | 2 | 3h |
| | | 3 | 5h |
| | | 4 | Same day delivery |
| | | 5 | Delivery within 2-4 days |
| 3. | Reputation/ ranking | 1 | High |
| | | 2 | Medium |
| | | 3 | Low |
| 4. | Apps (sending, tracking and tracing) | 2 | Yes/No |
| 5. | Apps (electronic delivery notification) | 2 | Yes/No |
| 6. | Personalization for delivery time window | 2 | Yes/No |
| 7. | Personalization for location of delivery | 1 | Home |
| | | 2 | Other (i.e. your car's trunk) |
| | | 3 | Pickup at a carrier's store |
| 8. | Payment method | 1 | On app/website (automatic) |
| | | 2 | By cash |
| 9. | Willingness to tip (this attribute is only for the questionnaire in the US) | 1 | No tip |
| | | 2 | $1 |
| | | 3 | $2 |
| | | 4 | $3 |

### Table 1 (b). Attributes to understand the couriers' behavior of selecting items

| q | Attribute meaning | $l_q$ | Level meaning |
|---|-------------------|-------|---------------|
| 1. | Profit (values of shipping cost in the US are outside the parenthesis, in Vietnam are inside the parenthesis) | 1 | $13 (45,000 VND) |
| | | 2 | $11 (35,000 VND) |
| | | 3 | $9 (25,000 VND) |
| | | 4 | $7 (15,000 VND) |
| 2. | Trip time (addition to travel time of the original trip) | 1 | +20 minutes |
| | | 2 | +40 minutes |
| | | 3 | +60 minutes |
| 3. | Compensation due to loss or damage | 1 | 80% price (20% less) |
| | | 2 | 100% price (regular price) |
| | | 3 | 120% price (20% more) |
| 4. | Weight - denoted as x. (Units in the US are pounds, in Vietnam are kg) | 1 | x ≤0.5 |
| | | 2 | 0.5 < x <1.5 |
| | | 3 | x ≥1.5 |
| 5. | Number of items needed to be picked up and delivered (this attribute is only for the questionnaire in Vietnam) | 1 | 1 |
| | | 2 | 2 |
| | | 3 | 3 or more |

To capture the courier's behavior (part IIB), the SP was designed to identify prospective crowd-shippers' behaviors. Each alternative includes attributes which may influence their decision to deliver a package and their ETP. The choice set consists of four alternatives, the attributes of which are shown in Table 1 (b). The respondents were also asked for their perceptions of the product/ item category to be shipped, which packages or goods they would prefer to deliver, and their concerns regarding crowd-shipper employment.

To generate an orthogonal design, we used IBM SPSS Statistics 22 (*21*). Then, we eliminated options based on several rules, such as removing the dominant options, separated options into blocks, and selected representative options from each block. We followed previously established SP survey design techniques (*22*)-(*24*).

## Survey implementation

Pilot surveys were conducted to improve the quality of the questions and assess the time required for survey completion. Two pilot surveys were implemented with Vietnam and US participants from various disciplines, ages, genders, and occupations. After conducting pilot surveys and the pre-test, questionnaires were modified for the final survey.

The surveys were conducted via multiple channels. Taking advantages of the Transportation Research Board 2017 conference and committee membership, the authors actively distributed flyers during these events. In addition, the surveys were emailed to students at various colleges, schools, universities, chapters, and organizations. The surveys were also advertised on social media (i.e., Facebook, LinkedIn, Reddit, and Craigslist). An additional source for the US survey was Amazon Mechanical's Turk.

## DESCRIPTIVE ANALYSIS

This 2017 study intends to understand the behavior of stakeholders in the crowd-shipping market and surveyed thousands of people to gain insights into their preferences, concerns, and hesitations regarding crowd-shipping services. Results were organized into different socio-demographic characteristics.

## Summary of the data collection

The survey was implemented in the US between February and April 2017, and in Vietnam in April 2017. Samples from 1,176 (US) and 907 (Vietnam) respondents were collected. The data was then cleaned to remove incomplete and inconsistent responses. The final data sets included 549 and 509 samples for the US and Vietnam surveys, respectively.

## A brief summary of socio-demographic characteristics of respondents

The age distribution of survey respondents are displayed in Figure 1. Respondents in Vietnam were 31% male and 69% female, while those for the US survey were 46% male and 54% female. Age and gender were nicely distributed in the US data set; hence, this data set can be utilized to obtain insights for follow-up research. Respondents in Vietnam, however, were mainly young female students, so the results were used to identify crowd-shipping trends only. Other socio-demographic characteristics are presented in Table 2.

## Requesters' experience and expectations

The requesters' experiences and expectations are summarized in Table 3 (a). Regarding delivery carriers, self-employed transporters were reported to be the most popular courier in Vietnam. Transporters were hired by about 75% of online retailers, and by over 32% of respondents who requested package delivery. USPS and UPS were the two main carriers for the US respondents who sent packages (48%) and merchants (42%), respectively. Surprisingly, those carriers (i.e. self-employed transporters, USPS, and UPS) were also reported as the worst delivery service providers.

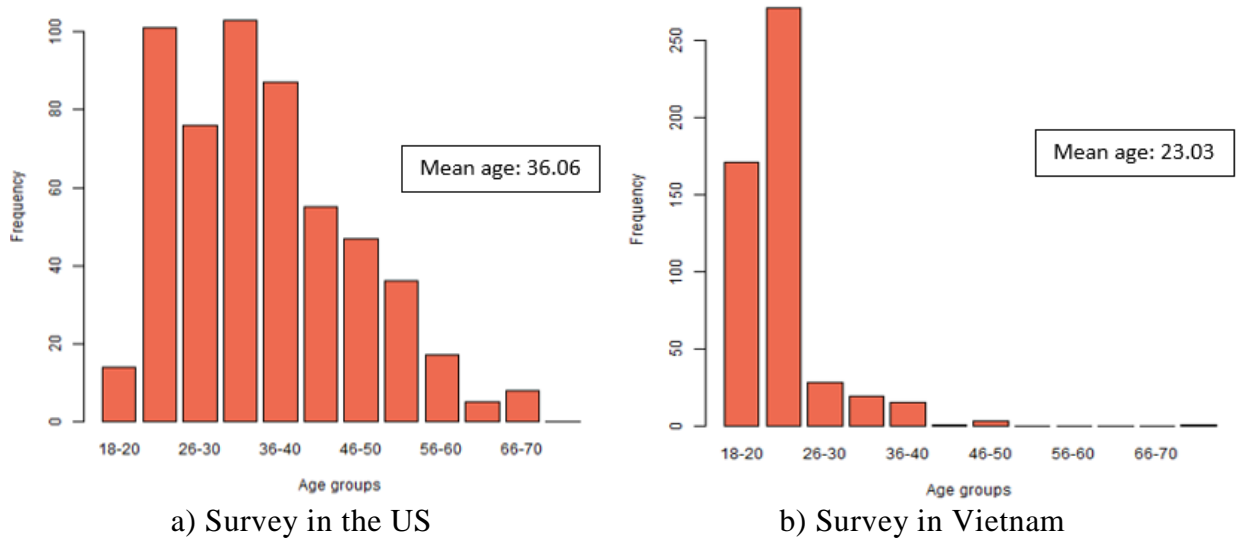

a) Survey in the US                              b) Survey in Vietnam

**Figure 1. Age distributions of respondents**

Requesters preferred to have "immediate" or "groceries" delivered by crowd-shippers, which involves higher delivery fees and shorter delivery times. For other products, however, they were more likely to use traditional carriers with lower delivery fees and longer delivery times (Table 3b). Various factors that requesters consider when selecting a courier are not yet clear, and require further research.

Respondents expected to have their packages delivered at different times on weekdays and weekends. The US respondents preferred to receive their packages in the late afternoon or early evening (3 p.m.-8 p.m.) on weekdays, and from 9 a.m. to 6 p.m. during the weekend. Vietnam respondents preferred to receive packages from 6 p.m. to 10 p.m. on weekdays, and from 9 a.m. to 10 p.m. during the weekend. This information is useful for crowdsourced delivery companies' operation (e.g. prepare crowd-shippers for high demand times, and prevent false delivery). In addition, we added drones as an option with other modes of transport to examine respondents' preference for courier modes. Interestingly, both the US and Vietnam respondents did not have any preference ("it does not matter"), 65% and 37%, respectively. Furthermore, Vietnam respondents liked to have their packages carried by motor bike (53%) or car (33%). 5% to 7% of respondents were in favor of using drones. Additional information is included in Table 3 (b).

Respondents were then asked for a stated preference question regarding their concerns once their package is delivered by a crowd-shipper. The major concern (around 85%) related to the condition of the packages: "without damage or not." Moreover, "delivery on time or not" was another concern of 67% and 46% of the US and Vietnam respondents, respectively.

The requesters' other behaviors, such as which groups of people (or which type of commodity) are more sensitive to price, and which factors influence requester selection of couriers, will be extensively studied in another paper.

**Note:** [*] values and distribution by percentage are statistics for indicator and other variables.
- (sign): the variable/option is not available.
Bold font: the largest portion(s).
^number: corresponding answer's option.

### Table 2. Socio-demographic characteristics

| Variables | Unit | Min/ Max or Values[*] | Mean/ Standard Deviation or Distribution[*] | | | |
|---|---|---|---|---|---|---|
| | | | Total samples | | WTW | |
| | | | US (n = 549) | VN (n = 509) | US (n = 430) | VN (n = 407) |
| Age. | Years old | (US) 19/ 68; (VN) 19/ 73 | 36.06/ 11.06 | 23.03/ 5.14 | 36.42/ 10.79 | 22.66/ 4.08 |
| Gender: Male/ Female. | NA | 1/ 2 | 45.50/ 54.50 | 31.00/ **69.00** | 47.20/ 52.80 | 31.40/ **68.60** |
| Race/ ethnicity: African American^1/ American Indian, Alaska Native^2/ Asian^3/ Caucasian^4/ Hispanic, non-white^5/ Hispanic, white^6/ Others^7/ I prefer not to answer^8. | NA | 1/8 | 4.60^1/ 3.60^2/ 17.70^3/ **60.80**^4/ 3.30^5/ 5.30^6/ 3.30^7/ 1.50^8 | - | 4.90^1/ 4.20^2/ 16.50^3/ **61.40**^4/ 3.00^5/ 5.80^6/ 2.80^7/ 1.40^8 | - |
| Marital status: Single/ Married/ Others. | NA | 1/3 | **45.00**/ **44.80**/ 10.20 | **89.60**/ 9.60/ 0.80 | **43.30**/ **47.20**/ 9.50 | **90.20**/ 8.80/ 1.00 |
| Number of children. | Numbers | 0/ 6 | 0.94/ 1.25 | 0.20/ 0.64 | 1.00/ 1.27 | 0.18/ 0.60 |
| Number of people living in your household are less than 18 years old. | Numbers | 0/ 6 | 0.84/ 1.23 | 1.00/ 1.20 | 0.90/ 1.23 | 1.00/ 1.19 |
| Number of people living in your household are from 18 to 64 years old. | Numbers | 0/ 6 | 1.59/ 1.23 | 3.06/ 1.27 | 1.62/ 1.23 | 3.07/ 1.26 |
| Number of people living in your household are 65 years old or older. | Numbers | 0/ 6 | 0.17/ 0.60 | 0.64/ 0.87 | 0.17/ 0.56 | 0.49/ 0.85 |
| Final academic degree: Some high school^1/ High school diploma^2/ Technical college degree^3/ College degree^4/ Post-graduate degree^5 / I prefer not to answer^6. | NA | 1/6 | 0.40^1/ 12.90^2/ 8.60^3/ **48.50**^4/ 29.00^5/ 0.70^6 | 0.20^1/ 3.90^2/ 2.40^3/ **84.50**^4/ 9.00^5/ -^6 | 0.50^1/ 13.00^2/ 8.10^3/ **49.80**^4/ 28.40^5/ 0.20^6 | 0.20^1/ 4.20^2/ 2.20^3/ **86.00**^4/ 7.40^5/-^6 |
| Employment status: Employed full time^1/ Employed part-time^2/ Student (RA/ TA) (US)^3 or funded by family (VN)^3/ Student (having scholarship/fellowship)^4/ Student (self-funded)^5/ Retired and looking for job^6/ Retired and not looking for job^7/ Unemployed (US) or self-employed (VN) and looking for job^8/ Unemployed (US) or self-employed (VN) and not looking for job^9/ Others^10/ I prefer not to | NA | 1/11 | **48.10**^1/ 16.00^2/ 9.50^3/ 5.50^4/ 6.20^5/ 0.40^6/ 2.20^7/ 4.60^8/ 4.40^9/ 2.90^10/ 0.40^11 | 13.20^1/ 5.30^2/ **49.10**^3/ 6.90^4/ 21.40^5/ 3.10^6/ 0.80^7/ 0.00^8/ 0.00^9/ 0.20^10/ -^11 | **49.80**^1/ 16.50^2/ 8.60^3/ 4.90^4/ 5.60^5/ 0.50^6/ 1.90^7/ 4.90^8/ 3.50^9/ 3.50^10/ 0.50^11 | 12.50^1/ 5.70^2/ **46.90**^3/ 6.40^4/ 23.60^5/ 3.90^6/ 0.70^7/ 0.00^8/ 0.00^9/ 0.20^10/ -^11 |

| Variables | Unit | Min/ Max or Values* | Mean/ Standard Deviation or Distribution* | | | |
|---|---|---|---|---|---|---|
| | | | Total samples | | WTW | |
| | | | US (n = 549) | VN (n = 509) | US (n = 430) | VN (n = 407) |
| answer^11. | | | | | | |
| Annual income ((US) 1,000$; (VN) million VND). | $/VND | 15/ 220 | 48.71/ 36.00 | 43.61/ 37.35 | 47.87/ 33.38 | 41.73/ 34.96 |
| Type of accommodation: Owned/ Mortgage/ Rented/ Others. | NA | 1/4 | 29.50/ 20.00/ **49.00**/ 1.50 | **41.30**/ -/ 14.30/ **44.40** | 31.20/ 21.40/ **46.30**/ 1.20 | **40.30**/ -/ 15.20/ **44.50** |
| Have car or motorbike operator license: No/ Yes. | NA | 0/ 1 | 12.60/ 87.40 | 35.40/ 64.60 | 11.40/ 88.60 | 35.10/ 64.90 |
| Ownership: (US) car; (VN) Motorbike: No/ Yes. | NA | 0/ 1 | 19.30/ 80.70 | 52.10/ 47.90 | 18.10/ 81.90 | 51.10/ 48.90 |
| Mode usually used for commute to work/ school: Walking^1/ Bike^2/ Motor^3/ Car^4/ Bus^5/ Others transit mode (e.g. subway)^6/ Others^7. | NA | 1/ 7 | 14.60^1/ 5.10^2/ 2.70^3/ **65.90**^4/ 8.60^5/ 2.20^6/ 0.90^7 | 15.30^1/ 11.60^2/ **40.10**^3/ 1.00^4/ 29.30^5/ -^6/ 2.80^7 | 13.70^1/ 4.90^2/ 3.30^3/ **66.50**^4/ 8.40^5/ 2.30^6/ 0.90^7 | 16.00^1/ 12.00^2/ 0.50^4/ 27.50^5/ -^6/ 2.70^7 |
| Mode do you usually use for other purpose: Walking^1/ Bike^2/ Motor^3/ Car^4/ Bus^5/ Others transit mode (e.g. subway)^6/ Others^7. | NA | 1/ 7 | 13.70^1/ 6.60^2/ 2.20^3/ **69.20**^4/ 6.40^5/ 1.30^6/ 0.70^7 | 1.00^1/ 9.80^2/ **50.10**^3/ 4.50^4/ 32.20^5/ -^6/ 2.40^7 | 13.00^1/ 7.90^2/ 2.80^3/ **68.80**^4/ 5.60^5/ 1.20^6/ 0.70^7 | 1.00^1/ 10.10^2/ **52.80**^3/ 3.40^4/ 30.50^5/ -^6/ 2.20^7 |
| Total access time to the closest transit station/ bus stop. | Minutes | (US) 0/ 32; (VN) 0/42 | 23.68/ 11.18 | 34.11/ 15.04 | 23.43/ 11.30 | 34.85/ 14.78 |
| Using smart phone: No/ Yes. | NA | 0/ 1 | 4.70/ 95.30 | 5.90/ 94.10 | 5.10/ 94.90 | 6.40/ 93.60 |
| Social media usages: Yes, frequently^1/ Yes, sometimes^2/ Yes, occasionally^3/ Yes, rarely^4/ No, not at all^5. | NA | 1/5 | **77.60**^1/ 10.40^2/ 5.30^3/ 2.70^5 | **88.80**^1/ 9.00^2/ 0.20^3/ 0.80^5 | **77.90**^1/ 10.20^2/ 5.10^3/ 4.00^4/ 2.80^5 | **89.90**^1/ 8.40^2/ 0.20^3/ 1.00^4/ 0.50^5 |
| What social media do you use: Facebook^1/ Twitter^2/ YouTube^3/ Reddit^4/ Tumblr^5/ Instagram^6/ Pinterest^7/ Vine^8/ Ask.fm^9/ Flickr^10/ Google+^11/ LinkedIn^12/ VK^13/ Meetup^14/ Others^15. | NA | 1/15 | **90.89**^1/ 43.89^2/ **70.86**^3/ 22.95^4/ 8.20^5/ 45.90^6/ 36.43^7/ 1.82^8/ 0.54^9/ 4.00^10/ 28.42^11/ 37.52^12/ 0.36^13/ 4.00^14/ 2.19^15 | **98.82**^1/ 6.67^2/ **66.40**^3/ -^4/ -^5/ 34.57^6/ -^7/ -^8/ -^9/ 1.77^10/ 45.00^11/ 2.55^12/ -^13/ -^14/ 5.70^15 | **90.93**^1/ 45.58^2/ **73.25**^3/ 24.19^4/ 8.84^5/ 46.05^6/ 36.51^7/ 1.86^8/ 0.70^9/ 3.72^10/ 30.00^11/ 37.67^12/ 0.46^13/ 3.72^14/ 1.86^15 | **98.53**^1/ 6.88^2/ **65.84**^3/ -^4/ -^5/ 32.67^6/ -^7/ -^8/ -^9/ 0.21^10/ 46.19^11/ 2.70^12/ -^13/ -^14/ 5.41^15 |
| Total number of social media uses. | Numbers | (US) 0/ 10; (VN) 0/8 | 4.00/ 2.05 | 2.61/ 1.28 | 4.07/ 2.08 | 2.60/ 1.27 |

## Table 3 (a). Courier selection behavior (RP)

| Variables | Unit | Min/ Max or Values* | Mean/ Standard Deviation or Distribution* | | | |
|---|---|---|---|---|---|---|
| | | | US (n1 = n11 + n12 = 549) | | VN ( n2 = n21 + n22 = 415) | |
| Experience: Sending package/ purchase online. | NA | 1/2 | 38.25/ 61.75 | | 23.86/ 76.14 | |
| **I, III: Sending package**<br>**II, IV: Purchase online** | | | **I.**<br>(n11 = 210) | **II.**<br>(n12 = 339) | **III.**<br>(n21 = 99) | **IV.**<br>(n22 = 316) |
| What have you sent to someone else/ you bought (multiple choices): Dry cleaning, fast foods, lunch, dinner, birthday cake, etc (immediate delivery)^1/ Groceries^2/ Beverage, dry foods^3/ Personal health, medicine^4/ Apparel^5/ Books, Music, Videos^6/ Consumer electronics^7/ Others^8. | NA | 1/8 | 6.48^1/ 6.78^2/ 5.30^3/ 12.97^4/ **38.93^5**/ **34.51^6**/ 25.68^7/ 22.41^8 | 10.95^1/ 14.28^2/ 10.95^3/ 23.80^4/ **34.76^5**/ 25.71^6/ **30.95^7**/ 17.61^8 | 13.13^1/ 20.20^2/ 19.19^3/ **32.32^4**/ **45.45^5**/ 15.15^6/ 21.21^7/ 14.14^8 | 12.02^1/ 3.80^2/ 7.91^3/ 32.91^4/ **62.97^5**/ 14.24^6/ 19.62^7/ 6.33^8 |
| (I, III) From where did you ship the item: Home/ Office/ Others;<br>(II) Which website/shop did you buy the item from: (II) Ebay^1/ Amazon^2/ ModCloth^3/ CololBlue^4/ Others^5;<br>(IV) Muachung/ Facebooks of retailers/ Others. | NA | (I, III, IV) 1/3; (II) 1/5 | **74.90**/ 13.30/ 11.80 | 6.20^1/ **71.00^2**/ 1.00^3/ 0.50^4/ 21.40^5 | **41.40**/ 34.30/ 24.20 | 3.20/ **75.00**/ 21.80 |
| (I, III) Approximate value of the item you requested to deliver;<br>(II, IV) How much did you pay for the item which you have bought it online. | (I, II) $; (III, IV) Thousand VND | (I) 5/6,000; (II) 4/3,000; (III) 1.2/50,000; (IV) 25/ 300,000 | 176.2/ 492.72 | 115.02/ 336.53 | 2,095.56/ 6,060.62 | 1,883.02/ 17,128.97 |
| Delivery carrier: (I, II) DHL^1/ UPS^2/ FedEx^3/ USPS^4/ By retail's personnel^5/ Others^6; (III, IV) DHL^1/ UPS^2/ FedEx^3/ EMS^4/ Giaohangnhanh^5/ Proship^6/ Hired transporter (or by retail's personnel) ^7/ Others^8. | NA | (I, II) 1/6; (III, IV) 1/8 | 6.20^1/ 28.60^2/ 15.30^3/ **47.50^4**/ 1.20^5/ 1.20^6 | 4.30^1/ **42.40^2**/ 20.50^3/ 26.20^4/ 5.20^5/ 1.40^6 | 6.10^1/ 0.00^2/ 1.00^3/ 17.20^4/ 16.20^5/ 1.00^6/ **32.30^7**/ **26.3^8** | 0.60^1/ 0.00^2/ 0.00^3/ 1.60^4/ 16.80^5/ 2.50^6/ **74.10^7**/ 4.40^8 |
| Payment method: (I, III) Paid for the courier by payment card (credit/ master/ etc) ^1/ Paid at store by payment card^2/ Paid for the courier by cash^3/ Paid at store by cash^4/ Paid online^5/ Others^6;<br>(II, IV) Paid for the courier once the package was delivered at | NA | (I) 1/5; (II) 1/4; (III) 1/6; (IV) 1/8 | 20.90^1/ 28.90^2/ 13.60^4/ **33.60^5**/ 2.90^6 | -^1/ 3.30^2/ -^3/ -^4/ 21.00^5/ **44.80^6**/ 31.00^7/ -^8 | 4.00^1/ 3.00^2/ 37.50^3/ **50.50^4**/ 3.00^5/ | **61.40^1**/7.90^2/ 3.20^3/ 3.20^5/ 0.30^6/ 22.80^7/ 0.30^8 |

| Variables | Unit | Min/ Max or Values[*] | Mean/ Standard Deviation or Distribution[*] | | | |
|---|---|---|---|---|---|---|
| | | | US ($n_1 = n_{11} + n_{12} = 549$) | | VN ($n_2 = n_{21} + n_{22} = 415$) | |
| home (or office/ shop/ etc) by cash^1/ Paid at store by cash^2/ Paid for the courier once the package was delivered at home (or office/ shop/ etc) by payment card^3/ Paid at store by payment card^4/ Paid online^5/ Free shipping (annual/prime member) ^6/ Free shipping (since one have paid over a certain amount for my purchase) ^7/ Others^8. | | | | | 2.00^6 | |
| Time for deliver. | Hours | (I) 0.75/23; (II) 0.5/ 19; (III) 0.5/12; (IV) 0.33/ 18 | 7.65/ 1.27 | 4.25/ 2.01 | 3.73/ 0.36 | 4.39/ 0.28 |
| Time for deliver. | Days | (I) 1/28; (II) 1/18; (III) 1/14; (IV) 1/30 | 3.22/ 0.09 | 3.46/ 0.13 | 2.75/ 0.24 | 2.64/ 0.13 |
| Satisfaction with the delivery time: No/ Yes. | NA | 0/1 | 5.90/**94.10** | 9.00/ **91.00** | 13.10/ **86.90** | 13.00/ **87.00** |
| Able to track and trace the item online: Yes, I could track via carrier's website (and/or app)/ Yes, I could track by calling courier's cellphone / No, I could not track it. | NA | (I, II) 1/2; (III, IV) 1/3 | **94.40**/ -/ 5.60 | **91.90**/ -/ 8.10 | 33.30/ 31.30/ 35.40 | 27.20/ 25.60/ **47.20** |
| Satisfaction with electronic delivery notification: Yes, I was satisfied with the service/ No, the service was not good/ No, they did not provide the service. | NA | 1/3 | **82.60**/ 3.80/ 13.60 | **84.30**/ 4.80/ 11.00 | **61.60**/ 10.10/ 28.30 | **68.00**/ 8.50/ 23.40 |
| Choose the pickup (I, III)/delivery (II, IV) time window: Yes, I could, and I used that service/ Yes, I could, but I did not use that service/ No, I could not. The carrier did not offer that service. | NA | 1/3 | 19.50/ 36.60/ **44.00** | 30.00/ 20.50/ **49.50** | **41.40**/ 17.20/ 41.40 | **56.60**/ 24.40/ 19.00 |
| Carriers offer pickup at home (I, III)/ provide convenient drop-off location (II, IV): Yes, but I have never used the service^1/ Yes, the service is excellent^2/ Yes, the service is good^3/ No, the service is bad^4/ I have no idea about the service^5. | NA | 1/5 | **33.90**^1/ 19.80^2/ 17.40^3/ 5.00^4/ 23.90^5 | 20.00^1/ **36.20**^2/ 24.80^3/ 1.90^4/ 17.10^5 | **28.30**^1/ 19.20^2/ **27.30**^3/ 9.10^4/ 16.20^5 | 26.90^1/ **30.70**^2/ **31.60**^3/ 0.90^4/ 9.80^5 |
| Did you tip the delivery person: No/ Yes. | NA | 0/1 | 92.90/ 7.10 | 97.60/ 2.40 | - | - |
| Number of times did: (I, II) you use the service of the carrier per year; (II, IV) you shop in the same website/shop per year. | Numbers | (I, II) 1/300; (III) 1/ 550; (IV) 1/500. | 18.41/ 35.92 | 19.23/ 27.98 | 25.05/ 73.87 | 7.07/ 31.01 |

**Table 3 (b). Courier selection behavior - Sending a package (SP)**

| Variables | Unit | Min/ Max or Values* | Mean/ Standard Deviation or Distribution* | |
|---|---|---|---|---|
| | | | **US (n = 1098)** | **VN (n = 1018)** |
| Courier selection (Dry cleaning, fast foods, lunch, dinner, birthday cake, etc (immediate delivery)): couriers 1 - 4 | NA | 1/4 | **31.40**/ **26.00**/ **24.80**/ 17.90 | **39.00**/ **25.30**/ **30.10**/ 5.60 |
| Courier selection (Groceries): couriers 1 - 4 | NA | 1/4 | **27.90**/ **26.70**/ 21.90/ 23.60 | **29.90**/ **33.00**/ **33.00**/ 4.10 |
| Courier selection (Beverage, dried foods): couriers 1 - 4 | NA | 1/4 | **26.00**/ **26.70**/ 19.30/ **28.10** | **33.50**/ 24.30/ 20.60/ 21.60 |
| Courier selection (Personal health, medicine): couriers 1 - 4 | NA | 1/4 | 23.10/ 21.40/ 20.20/ **35.20** | 30.70/ 20.20/ 13.50/ **35.60** |
| Courier selection (Apparels): couriers 1 - 4 | NA | 1/4 | 21.00/ 19.00/ 10.50/ **49.50** | 31.00/ 18.20/ 12.90/ **37.90** |
| Courier selection (Books, Music, Videos): couriers 1 - 4 | NA | 1/4 | 23.10/ 17.60/ 11.90/ **47.40** | 30.80/ 19.60/ 14.10/ **35.40** |
| Courier selection (Consumer electronics): couriers 1 - 4 | NA | 1/4 | 22.90/ 16.00/ 15.70/ **45.50** | 31.20/ 19.00/ 16.50/ **33.30** |
| Courier selection (Others): couriers 1 - 4 | NA | 1/4 | 24.30/ 16.00/ 11.90/ **47.90** | **32.50**/ 20.40/ 14.30/ **32.70** |
| When do you prefer to get your packages to be delivered to your house (multiple choices): Weekdays, 6-9AM^1/ Weekdays, 9AM-noon^2/ Weekdays, noon-3PM^3/ Weekdays, 3-6PM^4/ Weekdays, 6-8PM^5/ Weekdays, 8-10PM^6/ Weekdays, 10PM-6AM next day^7/ Weekend, 6-9AM^8/ Weekend, 9AM-noon^9/ Weekend, noon-3PM^10/ Weekend, 3-6PM^11/ Weekend, 6-8 PM^12/ Weekend, 8-10PM^13/ Weekend, 10PM-6AM next day^14/ I do not have any preference. Any time is ok^15 | NA | 1/15 | 10.00^1/ 16.67^2/ 19.52^3/ 30.47^4/ **35.23**^5/ 19.04^6/ 7.61^7/ 10.95^8/ **33.80**^9/ **38.09**^10/ **34.76**^11/ 20.95^12/ 12.85^13/ 5.71^14/ 24.76^15 | 10.61^1/ 15.32^2/ 11.39^3/ 12.57^4/ **24.16**^5/ 16.31^6/ 1.57^7/ 13.95^8/ **27.11**^9/ 15.32^10/ **20.63**^11/ **19.45**^12/ 10.21^13/ 5.70^14/ 14.15^15 |
| Your concerns once your package is delivered by a crowd-shipper (multiple choices): Deliver on time or not/ Without damage or not/ Others | NA | 1/3 | 67.14/ 84.76/ 8.57 | 46.17/ 85.46/ 5.89 |
| Preference on the mode that the courier chooses (multiple choices): Drone^1/ Walking^2/ Bike^3/ Motor^4/ Car^5/ Bus^6/ Others transit mode (i.e. subway) ^7/ I do not have any preference; it does not matter^8/ Others^9 | NA | 1/9 | 7.14^1/ 10.00^2/ 12.38^3/ 14.76^4/ 29.52^5/ 8.09^6/ 5.23^7/ **64.76**^8/ 2.38^9 | 5.10^1/ 2.35^2/ 3.92^3/ **52.84**^4/ **33.01**^5/ 1.96^6/ -^7/ **36.54**^8/ 1.57^9 |

**Prospective crowd-shippers' willingness to work and expectations**

When asked about respondents' experience delivering freight, one in two had not delivered for someone else in the Vietnam data set. The rate of respondents who had not delivered for someone else (not experienced) were about three times the rate of respondents who had (experienced) in the US data set (Table 4 (a)).

Respondents were asked whether they were willing to work as a crowd-shipper. About 78% and 80% of the US and Vietnam respondents, respectively, were interested in crowd-shipping employment. Of the US respondents who were willing to work and not willing to work, 70% and 91%, respectively, did not have prior experience with carrying freight. Those percentages were 46% and 72%, respectively, in the Vietnam data set. The information is featured in Figure 2.

- *Characteristics of respondents not willing to work as crowd-shippers*

Surprisingly, incentive was not the major reason for respondents who were not willing to work as crowd-shippers. About 43% and 29% of the US and Vietnam respondents, respectively, reported that they did not have the time and refused crowd-shipping work. Other respondents in the US (37%) and Vietnam (40%) were simply not interested in crowd-shipping.

The mean incomes of the US and Vietnam respondents who were not willing to work were $51,700 and 51.1 million VND (i.e. $2,250). These mean incomes are higher than those of all respondents in the US and Vietnam data sets, which are approximately $48,700 and 47.9 million VND (i.e. $2,107), respectively. These statistics are expected; those respondents with higher incomes are less likely to be interested in working as crowd-shippers.

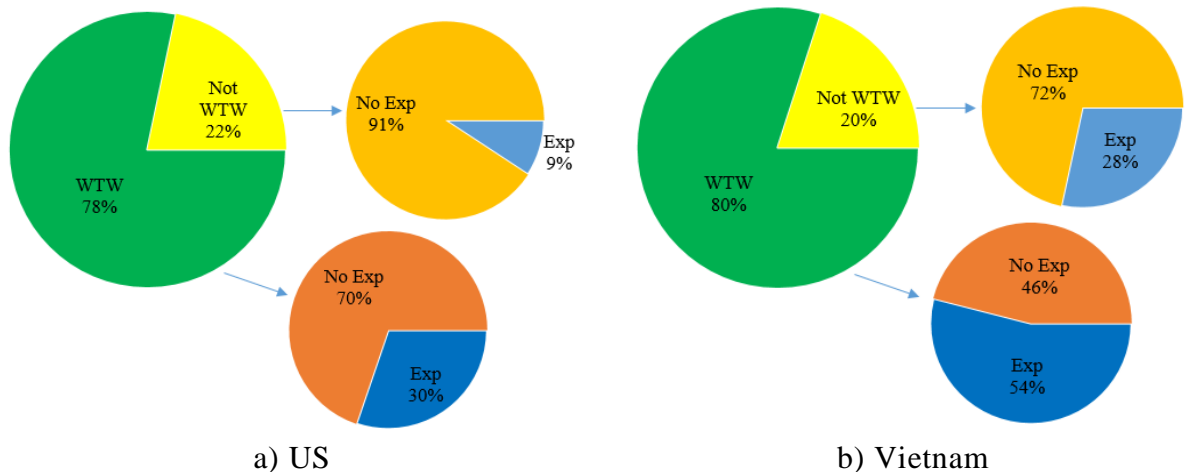

a) US                                                             b) Vietnam

**Figure 2. Descriptive statistics of the willingness to work as a crowd shipper**

- *Who is willing to work as a crowd-shipper?*

The socio-demographic characteristics of potential crowd-shippers are summarized in Table 2. Regarding the US data set, the average age of potential crowd-shippers was 36.42 years old, while the gender distribution was 47% and 53% male and female, respectively. There were just over 61% and 17% of respondents were Caucasian and Asian, respectively. In addition, 43% potential crowd-shippers were single and 47% were married. Around 78% had obtained or expected to earn a college degree or higher, and slightly over 21% held high school degrees. Approximately 50% and 17% of potential crowd-shippers were employed full time or part-time, respectively. Furthermore, potential driver partners earned $47,870 per year and had 1 child on average. A car was the main mode of respondents' transportation (approximately 70%).

The main social media outlets used by respondents in both data sets were Facebook (91% - 99%), YouTube (66% - 73%), Google+ (30% - 46%), and Instagram (33% - 46%). Moreover, the US respondents also used Twitter (46%), Pinterest (37%), and LinkedIn (38%). On average, the US respondents used more social media than those in Vietnam, with average of 4.09 outlets compared to 2.6. In addition, about 95% of respondents used a smartphone. Therefore, social media is a potential channel from crowd-shipping promotion and driver partner recruitment.

● *Perceptions of respondents willing to work as crowd-shippers* (Table 4 (a) and (b))
There were two major reasons for Vietnam respondents to work as crowd-shippers. "To be their own boss" was the main motivation (56%), for which 70% were females and 30% are males. On the other hand, 33% said "To earn money while looking for a full-time job," for which 66% and 34% were females and males, respectively. In addition, Vietnam respondents said that they would spend income from crowd-shipping mainly on monthly bills (33%) or "Treat for yourself/family" (49%).

Compared to Vietnam respondents, the US respondents were more likely to work as crowd-shippers for different trip purposes. The US respondents were willing to deliver packages during their commutes (70%), leisure trips (50%), and in their free time (70%). However, Vietnam respondents were more likely to work as crowd-shippers in their free time (85%) and during their commutes (52%). Potential crowd-shippers were likely to work during weekday evenings and weekend afternoons, times which highly matched with requesters' desired delivery times. This makes it much easier for crowd-shipping companies to pair requests and crowd-shippers.
The majority of respondents in both Vietnam (79%) and the US (87%) were willing to deliver any packages or goods, as long as they get paid. In addition, over 30% of the US respondents would also prefer to deliver to people they know (i.e., friends, colleagues, relatives, and neighbors). As such, crowd-shipping can be linked to driver partners' social networks to increase system demand. Furthermore, about 60% of the US respondents did not have any preference for the type of item they deliver; Vietnam respondents preferred to transport personal health/medicine (60%), apparel (70%), and books/music/videos (50%). Over 82% and 93% of the US potential crowd-shippers, and 66% and 77% of Vietnam potential crowd-shippers, expressed concerns about transporting "Hazardous materials/dangerous items" and "Illegal substances/products," respectively.

**Table 4 (a). Willingness to join a crowd-shipping system (RP and perceptions)**

| Variables | Unit | Min/ Max or Values[*] | Mean/ Standard Deviation or Distribution[*] | |
|---|---|---|---|---|
| | | | US (n = 430) | VN (n = 407) |
| Experience with transporting packages for somebody: No/ Yes. | NA | 0/ 1 | **74.32**/ 25.68 (n=549) | 51.08/ 48.92 (n=509) |
| WTW as a crowd-shipper: No/ Yes. | NA | 0/ 1 | 21.68/ **78.32** (n=549) | 20.04/ **79.96** (n=509) |
| Motivates you to work as a crowd-shipper (multiple choices): Maintaining a steady income because your other sources of income are unstable or unpredictable^1/ Earning more income^2/ To have more flexibility in my schedule and balance work with my life and family^3/ To be your own boss^4/ To earn money while looking for a full-time job^5/ Others^6. | NA | 1/6 | - | 16.71^1/    18.92^2/ 19.90^3/    **56.02**^4/ **33.42**^5/ 3.93^6 |
| Spend income from working as a crowd-shipper on (multiple choices): Monthly bills^1/ House renting^2/ Treat for yourself, family^3/ Expenses for your children^4/ Saving for emergencies^5/ Paying medical bills^6/ Student loans^7/ Saving for a big purchase^8/ Saving for retirement^9/ Others^10. | NA | 1/10 | - | **32.80**^1/    21.61^2/ **48.72**^3/    12.77^4/ 13.16^5/    23.57^6/ 5.50^7/      6.28^8/ 4.52^9/ 6.29^10 |
| Situations would you like to be a crowd-shipper (multiple choices): During my commute/ During my leisure trips/ In my free time/ Others. | NA | 1/ 4 | **70.00**/ 50.00/ **70.23**/ 1.62 | **51.84**/ 10.56/ **84.77**/ 0.49 |
| Total numbers of situations where you would be a crowd-shipper. | NA | 1/ 4 | 1.50/ 1.05 | 1.47/ 0.59 |
| When would you like to ship the freight (multiple choices): Weekdays, morning time^1/ Weekdays, afternoon time^2/ Weekdays, evening time^3/ Weekend, morning time^4/ Weekend, afternoon time^5/ Weekend, evening time^6/ Others^7. | NA | 1/ 7 | **42.79**^1/    38.60^2/ **44.88**^3/    40.00^4/ **44.65**^5/    34.18^6/ 2.79^7 | 16.95^1/    18.43^2/ 24.32^3/    **43.24**^4/ **51.84**^5/    29.73^6/ 4.42^7 |
| Total time slots you like to ship the freight. | Numbers | 1/ 7 | 1.94/ 1.61 | 1.89/ 0.99 |
| Maximum diversion (as a percent of distance). | % | 3/ 100 | 31.24/ 19.22 | 43.76/ 23.03 |
| Maximum distance (base 5 miles (US); base 8 kilometers (VN)). | Miles/ Km | 0/ 50 | 12.16/ 10.66 | 13.23/ 10.73 |
| Maximum diversion (in time) (base 20 minutes (US); base 30 minutes (VN)). | Minutes | 0/ 100 | 23.40/ 117.51 | 33.15/ 19.51 |

| Variables | Unit | Min/ Max or Values* | Mean/ Standard Deviation or Distribution* | |
|---|---|---|---|---|
| | | | US (n = 430) | VN (n = 407) |
| ETP as a crowd-shipper (base $15 (US); base 30,000 VND (VN)). | $/ 1,000 VND | (US) 0/ 30; (VN) 0/ 40 | 11.70/ 4.59 | 25.91/ 7.81 |
| Why you may NOT deliver freight for somebody else: The incentive (money paid) is not high enough/ I do not have time/ I do not like to do it/ Others. | NA | 1/ 4 | 9.20/ **42.90**/ 37.00/ 10.90. (n = 119) | 9.8/ 28.40/ **40.20**/ 21.60. (n = 104) |

**Table 4 (b). Willingness to join a crowd-shipping system (SP and preference)**

| Variables | Unit | Min/ Max or Values* | Mean/ Standard Deviation or Distribution* | |
|---|---|---|---|---|
| | | | US (n = 430) | VN (n = 407) |
| Selecting packages: items 1 – 4. | NA | 1/ 4 | 21.60/ 37.30/ 24.90/ 16.20 | 25.70/ 29.10/ 27.80/ 17.4 |
| Preference for the item to be shipped: Dry cleaning, fast foods, lunch, dinner, birthday cake, etc (immediate delivery)^1/ Groceries^2/ Beverage, dry foods^3/ Personal health, medicine^4/ Apparel^5/ Books, Music, Videos^6/ Consumer electronics^7/ No preference-do not care^8. | NA | 1/ 8 | 23.72^1/ 25.58^2/ 26.04^3/ 36.51^4/ 48.13^5/ 46.74^6/ 33.72^7/ **60.00**^8 | 20.39^1/ 10.07^2/ 33.90^3/ **59.71**^4/ **69.53**^5/ 49.89^6/ 16.71^7/ 11.06^8 |
| Whose packages or goods would you prefer to deliver: Your close friends, close colleagues^1/ Your friends, colleagues^2/ Your relatives^3/ Your neighbors^4/ Whosoever, I do not care once I get paid^5/ Others^6. | NA | 1/ 6 | 42.32^1/ 36.74^2/ 39.76^3/ 32.55^4/ **87.44**^5/ 1.16^6 | 20.39^1/ 23.09^2/ 21.13^3/ 14.99^4/ **79.11**^5/ 0.98^6 |
| What would be your concerns if you choose to work as a crowd-shipper: Hazardous materials, dangerous items^1/ Illegal substances, products^2/ Insurance if something bad happens^3/ Person is not at home^4/ Others^5. | NA | 1/ 5 | **82.09**^1/ **93.02**^2/ 2.79^3/ -^4/ 11.62^5 | **66.09**^1/ **76.62**^2/ 36.11^3/ **71.25**^4/ 1.71^5 |

Men and women had different preferences for the crowd shipping market. Women were willing to divert their routes for longer distances and more time than men. Moreover, women expected to be paid higher than men based on our samples. The Vietnam respondents tended to prefer to deviate shorter distances and times (in terms of percentages) than the US respondents. However, the former expected to be compensated higher than the latter. These factor comparisons between men and woman are summarized in Table 5.

## DISCUSSIONS AND RECOMMENDATIONS

The available crowd-shipping services in the logistics market provide additional options for customers. However, the following questions remain: what should crowd-shipping firms address to develop sustainable and lucrative business models, satisfy customers, and retain driver-partners? What should policymakers and local government do to achieve improved mobility, safety, and environmental sustainability? This section will further discuss our survey data insights that address these important questions.

- About 80% of respondents in our data set was willing to work as crowd-shippers. This statistic could have a significant impact on logistics carriers and society as a whole. Traditional logistics companies can reduce operation costs—for example, double-parking ticket costs—by outsourcing package delivery to crowd-shippers. Society can potentially benefit as well from the decreased environmental impact of less delivery vehicles. In order to attract people to work for the system, it is crucial to identify the motivations to work as crowd-shippers, as well as why people do not want to work for the system. Prospective crowd-shippers seem to be motivated by *economic* factors (e.g. "to earn money while looking for a full-time job", "earning more income", "maintaining a steady income"), while people refuse to work as crowd-shippers mostly due to *non-economic* reasons (e.g. "I do not have time", "I would not like to do it") (Table 4 (a)). Additional detailed analysis should be conducted to identify other factors related to these decisions.
- Travelers are a heterogeneous group. Some may be willing to transport freight but do not know where to find it. Some may transport freight for a sufficient incentive. Others may only transport freight for people who they know. Some may never work in the system due to their personal preferences or constraints. As can be seen in Table 5, age, number of children, and car ownership are significant factors in the ANOVA analysis of potential male and female US crowd-shippers. Additional detailed analysis should be conducted to better understand the behavior of travelers that are essential to crowd-shipping business strategies (e.g. recruit potential driver partners and develop compensation schemes). For example, insights derived from the results may identify a target group of applicants; for example, people with children, full-time workers with lower incomes, or part-time workers. Targeting a specific demographic will save time and costs for up-and-coming crowd-shipping businesses.
- The compensation scheme is a crucial factor to recruit and maintain occasional drivers in the crowd-shipping system. The ANOVA test shows the payment expectations of potential male and female crowd-shippers in the US is a significant factor (Table 5). Female crowd-shippers ETP paid higher than male driver partners. Further analysis is necessary to investigate the factors

**Table 5. Summary of potential crowd-shippers' behavior and socio-demographic characteristics by gender**

| Variables | US | | | VN | | |
|---|---|---|---|---|---|---|
| | Mean (percentage compared to base) or mean or percentage | | ANOVA test | Mean (percentage compared to base) or mean or percentage | | ANOVA test |
| | Male (47%) | Female (53%) | | Male (31%) | Female (69%) | |
| Maximum distance (base 5 miles (US)/ 8 kilometers (VN)) | 11.69 (233.8%) | 12.57 (251.4%) | N | 12.40 (155.0%) | 13.62 (170.3%) | Y |
| Maximum diversion (time) (base 20 minutes (US)/ 30 minutes (VN)) | 22.55 (112.8%) | 24.16 (120.8%) | N | 31.83 (106.1%) | 33.76 (112.5%) | N |
| ETP as a crowd-shipper (base $15 (US)/ 30,000 VND (VN)) | 11.42 (76.1%) | 11.95 (79.7%) | Y | 25.61 (85.4%) | 26.03 (86.8%) | N |
| In which situation: During my commute; During my leisure trips; In my free time | **70%**; 49%; **65%** | **71%**; 51%; **75%** | - | 46%; 11%; **93%** | 55%; 11%; **81%** | - |
| When would you like to ship the freight? Weekdays: morning time^1, afternoon time^2, evening time^3; Weekend: morning time^4, afternoon time^5, evening time^6 | 36%^1; 35%^2; **47%**^3; **42%**^4; **45%**^5; 37%^6 | **49%**^1; **42%**^2; **43%**^3; 39%^4; **44%**^5; 32%^6 | - | 16%^1; 21%^2; 31%^3; **45%**^4; **50%**^5; 35%^6 | 18%^1; 17%^2; 22%^3; **43%**^4; **53%**^5; 27%^6 | - |
| Age (years old) | 33.71 | 38.78 | Y | - | - | - |
| Marital status: single; married | 51%; 43% | 37%; 52% | - | - | - | - |
| Number of children | 0.69 | 1.27 | Y | - | - | - |
| Having college degree or above | 83% | 74% | - | - | - | - |
| Working full time; part time | **52%**; 13% | **48%**; 20% | - | - | - | - |
| Income | 46.82 | 48.82 | N | - | - | - |
| Car ownership | 79% | 84% | Y | - | - | - |
| Number of social media usages | 3.99 | 4.15 | N | 2.65 | 2.58 | N |
| Use social media daily | 77% | 78% | - | 89% | 90% | - |

ANOVA test: Y = significant, N = not significant.
Bold font: the largest portion(s).

relating to how much crowd-shippers ETP for the time of the day and day of the week as well as the sensitivity of the incentive to the diversion in travel time or travel distance. These insights are valuable for business model development (e.g. to determine a delivery price that is attractive for requesters but also compensates driver partners to participate in the system).

• Some respondents who do not own a car or smartphone are willing to work as crowd-shippers (Table 2). Therefore, providing a low-interest car loan or free smartphone can be a potential solution to attract those respondents to become crowd-shippers. For those who use other modes of transport and are willing to work as crowd-shippers, matching the request to the public transport schedule (e.g. metro or bus) on the same platform would facilitate increased participation.

• Individual marketing plays an important role in ridesharing service implementation (*25*); accordingly, it may be a successful promotion strategy for crowd-sourced delivery services. Our two data sets illustrate that most respondents frequently use multiple social media platforms (Table 2 and Table 3a). As such, there is a high potential to attract participation and shippers via social media.

• The performance measurement is a key feature of the crowd-shipping system. Since no operational data from crowd-shipping companies exists at this time, using travel survey data to forecast the potential supply is essential to industry performance. Simulation scenarios could feature the acceptance rate of travelers who may be willing to work as crowd-shippers. The corresponding benefits to society, such as congestion, safety, and pollution, should be estimated as well. Performance measurements of the crowdsourced delivery system, driver partners' incentive preferences, and long-term traveler behaviors are all important metrics that should be understood more clearly.

• Local governments can ensure the effective implementation of crowd-shipping services by offering incentives to system stakeholders. Possible incentive examples include tax cuts on the income earned from crowd-shipping services, free priority parking at designated locations, free congestion pricing fee, and driver partners use of high-occupancy vehicle lanes. Therefore, the effectiveness of the system can be measured by various incentives. Government then should identify which incentives or incentive packages are the most attractive to travelers and the most effective to modify travelers' behaviors. Moreover, the data (e.g. speed and travel time) collected from crowd-shipping firms can be integrated into citywide management apps to improve transportation operations.

## CONCLUSIONS

Crowd-shipping is a relatively new research topic, and various aspects of the field are worthy of transportation experts' exploration. We have examined the current and future behaviors of requesters and potential crowd-shippers given the availability of crowdsourced delivery services in the logistics market. The contributions of this study are summarized as follows:

• Our work is focused on designing questionnaires and conducting surveys in Vietnam and the US to investigate behavioral issues on both the demand and supply sides of crowd-shipping in the two countries. The questionnaires included RP and SP questions, which will be used in combined RP-SP models in future research. The SP questions were carefully designed to maximize the reliability, validity, and generalizability of the data.

• We have collected data on current shipping and purchasing delivery-related behaviors. A brief description has been conducted to understand respondents' tastes and satisfaction levels regarding delivery time, track and trace, electronic delivery notification, pickup/drop-off time window, and pickup/drop-off location of the most recent shipping and delivery activities. Most respondents were satisfied with the services, but those services still require improvement for a seamless experience and maximized customer satisfaction.

• Data on the awareness of SP requesters and potential crowd-shippers has also been collected. This study features descriptive analysis, but further investigation is needed to reveal the behaviors of requesters and potential crowd-shippers under the assumption of crowd-shipping service availability in the logistics market.

Directions for further studies and analysis include:

• Target features of crowd-shipping platforms related to cost effectiveness and convenience should be examined. The features include, but are not limited to, tracking and tracing the item online, electronic delivery notification and communications, and personalization of delivery time window and delivery location. These insights are crucial for crowd-shipping business operations such as request-courier matching and dynamic routing.

• Behavior differences by geographic location can be achieved for a bigger survey size or area. Then, we can build models with spatial explanatory variables like land-use, which includes the public transport density, living costs, intensity of commercial areas, developed areas, etc. Those independent variables may provide additional insights into the models as well as support for crowd-shipping companies and local governments implementing the service.

• Questionnaires in this research were intentionally designed for first- and last-mile deliveries. However, this knowledge can be adapted to conduct studies for the middle- and long-haul delivery services.

• Conduct a survey with current driver partners who are working for crowd-shipping companies. Insights derived from this data set can be used to validate findings in the literature.

• The impact of crowd-shipping is not yet clear, in particular its carbon footprint. Therefore, researchers should also investigate these issues. The successful implementation of crowd-shipping service are likely fueled by positive environmental impacts worth measuring.

In conclusion, this research is an important milestone in understanding the crowd-shipping market and provides various useful insights based on data from two countries. However, further research is necessary to get a fuller picture of various micro level details of the crowd-shipping market.

## ACKNOWLEDGMENT

Authors want to thanks our colleagues at the University of Transport and Communications, and at the Interdisciplinary Transportation Modeling and Analytics Lab at Purdue University for their support and advertising the survey. We also want to thanks the people who participated in the survey in Vietnam and the US. In addition, we are grateful for the comments and suggestions on the earlier version of this manuscript from Professor. Tom Van Woensel and Professor. Luuk Veelenturk of the Eindhoven University of Technology, the Netherland.